# Web-Based Learning and Training for Virtual Metrology Lab

Fahad Al-Zahrani

Computer Science Department, Faculty of Engineering, Albaha University, 65451 Albaha, KSA

**Abstract**— The use of World Web Wide for distance education has received increasing attention over the past decades. The real challenge of adapting this technology for engineering education and training is to facilitate the laboratory experiments via Internet. In the sciences, measurement plays an important role. The accuracy of the measurement, as well as the units, help scientists to better understand phenomena occurring in nature. This paper introduces Metrology educators to the use and adoption of Java-applets in order to create virtual, online Metrology laboratories for students. These techniques have been used to successfully form a laboratory course which augments the more conventional lectures in concepts of Metrology course at Faculty of Engineering, Albaha University, KSA. Improvements of the package are still undergoing to incorporate Web-based technologies (Internet home page, HTML, Java programming etc…). This Web-based education and training has been successfully class-tested within an undergraduate preliminary year engineering course and students reported a positive experience with its use. The use of these labs should be self-explanatory and their reliable operation has been thoroughly tested.

**Index Terms**— Web-based education, Education in Metrology, Virtual laboratory, Software engineering, Java, HTML

## 1 INTRODUCTION

Computer applications are used in many central sectors of our society: health, bank, security, etc. Of course, education is not an exception, and it introduces unique technical, managerial and most importantly pedagogical issues. These unique features make computers and education a field of study on its own [1]. The history of software engineering education was traced by Mead [2] which focused on some of the key players. Web-based education, Java and Shockwave applets, Web-based experiments, educational software, Web-based e-learning and evaluation will be reviewed in the following paragraphs.

Sethi introduced physics educators to the use and adoption of Java- and Shockwave-based applets in order to create virtual, online physics laboratories for their students [3]. These techniques have been used to successfully form a Laboratory course which augments the more conventional lectures in a concept of physics course. The laboratories have been instructor-led but are sufficiently self-contained to become part of a virtual classroom offering. Sethi and Antcliffe [4], described a set of physics experiments that use selected Java and Shockwave applets which form a Laboratory course that augments the more conventional lectures in a Concepts of Physics course. An educational software tool–editor that will as primary goal to create Java applications and applets was created by Fetaji [5] and at the same time to provide mechanism for the user to compile his source code without leaving the editor. His goal is also set to provide a complete e-learning environment for the course object oriented programming in Java language.

Pniower et al. [6] reported on two Web-based experiments operating since spring 1998: a Michelson interferometer and a laser diode characterization experiments. Descriptions include benchtop optical and electronic experimental hardware, LabVIEW software tools for hardware interfacing, HTML Web interfacing tools, and the video link setup.

On the other hands, the development of educational software was reported by Syropoulos et al. [7] to facilitate on-demand availability of traditional hydraulic programs written in any high-level computer language through the Web. This software has been successfully class-tested within an undergraduate civil engineering course and students reported a positive experience with its use. Iqbal and Siddiqui [8] discussed the software and hardware requirements of virtual courses/laboratories for electrical engineering courses to provide interactive environment for designing and conducting classes/experiments. This example shows how similar laboratories can be implemented for other courses like chemical engineering, mechanical engineering and system engineering etc.

The impact of Intercollege's Web-Based Teaching and Learning Environment which allows lecturers to develop Web-based educational material using Web-based educational software - WebCT (off the-shelf software) and InterLearning (in-house developed software) was presented by Pouyioutas et al. [9]. Aichouni and Al Nais [10], suggested that computer based education approach can be used as a learning and a training tools to demonstrate statistical quality control concepts and their industrial applications for engineering students. Active learning is an important approach to prepare engineers, especially manufacturing engineers. Sirinterlikci [11] presented various activities to engage students more actively in their manufacturing engineering education.

In addition, an introduction of modern information and communication technologies in education which focused on development of the multimedia tool for education in measurement and metrology was described by Halaj et al [12]. The information systems design theory (ISDT) has, over the last few years, produced a Web-based education information system that is more inclusive, flexible and is more closely integrated with the needs of its host organization. Jones et al. [13], introduced the origins of ISDT concept and described one ISDT for Web-based education. Milentijevic et al. [14] dealt with the development of a generalized model for version control systems application as a support in a range of project-based learning methods. The suggested model encompasses a wide range of different project-based learning approaches by assigning a supervisory role either to instructor or students in different project stages. Possible implementations of different project-based learning approaches on the proposed model are demonstrated by setting the model parameters.

Evaluation is an important component of developing educational software. Ideally, such evaluation quantifies and qualifies the effects of a new educational intervention on the learning process and outcomes. Silvast et al. [15] collected a logging data of the usage of a Web-based educational software for solving exercises on algorithms and data structures and assessed students' solutions automatically. Analysis of the logging data can be an important method for evaluating the efficacy of the software in improving students' learning. This opens up interesting research questions for understandings student behavior better, and to improve the education methods in the future. Schleyer and Johnson [16] used several published examples to illustrate different evaluation methods. They stated that readers are encouraged to contemplate a wide range of evaluation study designs and explore increasingly complex questions when evaluating educational software.

In order to engage the Net generation with this topic, Janzen and Ryoo [17] proposed a development and population of a community-driven Web database containing summaries of empirical software engineering studies. Motivations, student and instructor-developed prototypes, and assessments of the resulting artifacts are discussed. In fact, 30% more of the respondents found the student-written summaries to be ''very useful''. In order to improve knowledge acquisition during experience reuse, Ras and Rech [18] presented an approach based on Web 2.0 technologies that generates so-called learning spaces. This approach automatically enriches experiences with additional learning content and contextual information. To evaluate this approach, a controlled experiment was conducted, which showed a statistically significant improvement for knowledge acquisition of 204% compared to conventional experience descriptions. From a technical perspective, the approach provides a good basis for future applications that support learning at the workplace in academia and industry for the Next Generation.

This paper aims to adopt Java-applets in order to create virtual, online Metrology laboratories for students and to build a Web-based education and training for Metrology and measurement.

## 2 WEB-BASED EDUCATION AND TRAINING

Web-based educational experiments allow remote users to conduct laboratory explorations using metrological experimental apparatuses in real time over the World Wide Web (WWW). Web-based experimentation is evolving rapidly and offers students convenient and repeated access to limited laboratory resources. The immediacy and accessibility of Web-based experiments can also assist new student outreach and faculty teaching effectiveness [6]. WWW is a popular and useful instructional tool for a many reasons. It is easily accessible, supports flexible storage and display options and provides a simple yet powerful publishing format to incorporate multiple media elements. The WWW provides an excellent platform for developing, organizing, and spreading variety of resources, including class notes and outlines, long textual resources that resemble traditional textbooks, interactive nonlinear tutorials, student questions and comments, and even simulations of individual class sessions. It also allows instructors to prioritize resources and students to reorganize the resources in the way that fits them best. Many universities and colleges have utilized the WWW for developing distance-learning education courses [9, 19, 20]. Figure 1 show the Web-based education and its different areas of interest.

Velázquez-Iturbide and García-Peñalvo classified a nine articles into three groups, namely e-learning and Web-based educational software, user-centered educational software, and visualization tools [1], Fig 2. The first group deals with e-learning and Web-based educational software. e-Learning and its different blended-learning variants are nowadays ubiquitous, at least at the university level. The second group is concerned with user issues. There are many different issues in and forms of user-centered educational software which are: user motivation, automatic assessment and correctors, interactive tutors, accessibility, and support to learning styles. The third group focuses on a particular class of software technology, namely visualization. The aim of visualization can be stated as making visible what is hidden, by using graphical representations. These articles represent two kinds of visualization, namely information visualization and software visualization. Although the latter can be considered a particular case of the former, their research interests and communities are currently different.

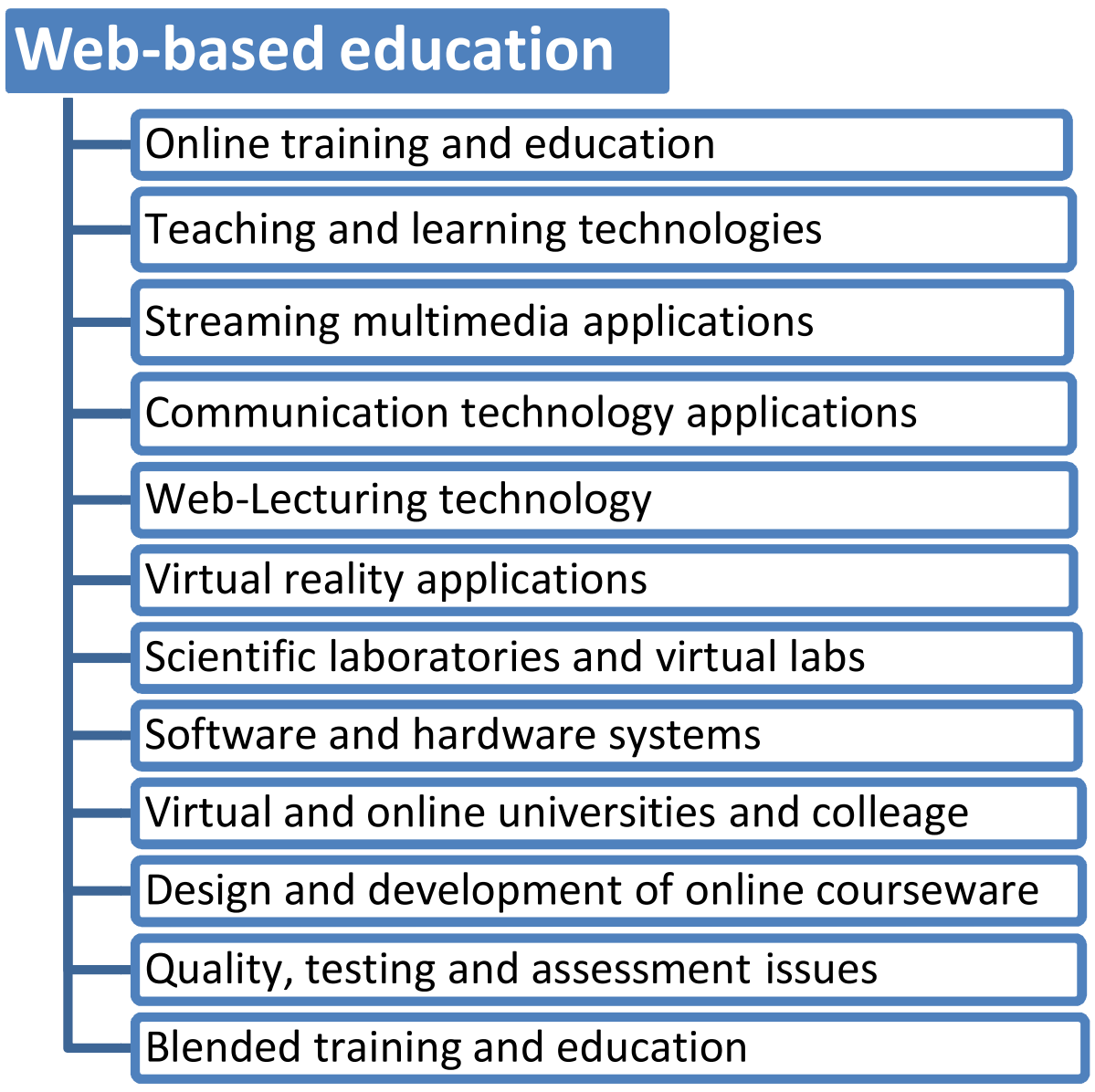

*Fig. 1. Classification of Web-based education.*

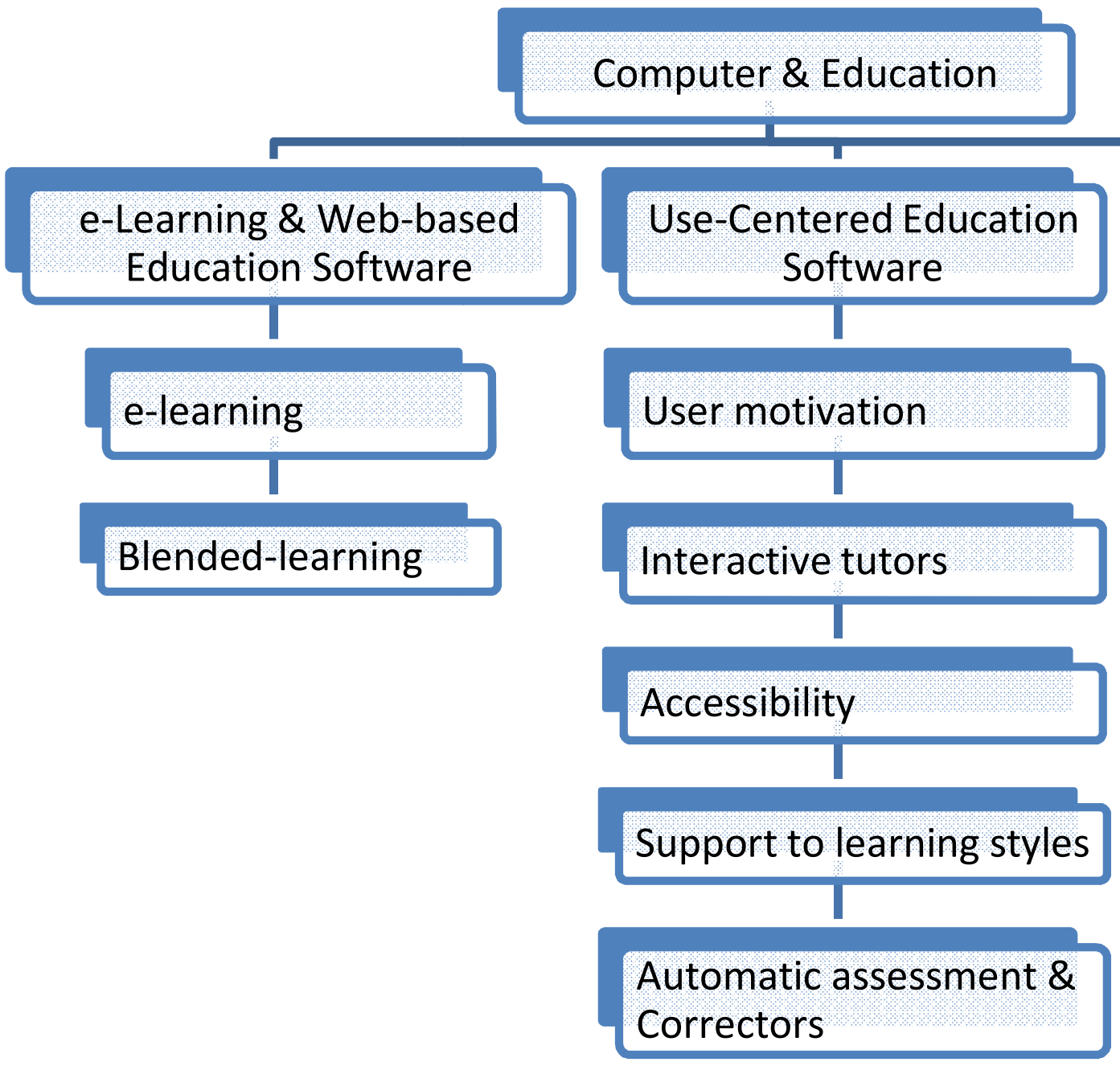

Fig. 2. Computer and Education.

The history of software engineering education was traced and what has been accomplished in degree programs and curricula, conferences and working groups, professionalism, certification, and industry–university collaboration was highlighted [2, 21]. Figure 3 summarizes the history of software engineering education since past to present.

**Past**

- **1968** "Software engineering" term first appeared at the NATO Software Engineering Conference.
- **1970s** Software engineering education appeared at IBM.
- In the late **1970s**, master's degree programs in software engineering started to emerge.
- A series of courses entitled Structured Programming Workshop (SPW), Structured Design Workshop (SDW), and eventually Advanced Design Workshop (ADW) were offered.
- **1984** Fairley wrote one of the first books on software engineering concepts and went on to become a leader in software engineering education.
- **1986/1987** The first Conference on Software Engineering Education (CSEE) took place.
- **1991** The first class of Multimedia Study Environment (MSE) graduate.
- In the early **1990s**, the CSEE became the catalyst for a newsletter, the Forum for the Advancement of Software Engineering Education (FASE).
- **1993** A joint committee was formed by Association for Computing Machinery (ACM) and Institute of Electrical and Electronics Engineers (IEEE) Computer Society to promote software engineering as a profession.
- **1995** A Working Group for Software Engineering Education and Training (WGSEET) was formed.
- **1995** Undergraduate degree programs in software engineering education started to emerge, and in the US a merger of the accrediting bodies, Computer Sciences Accreditation Board (CSAB) and Accreditation Board for Engineering Technologies (ABET), made it possible for them to be accredited.
- **1995** Many universities established industry advisory boards so that their degree programs could remain relevant to their industry partners. These collaborations took many forms—joint research, regular work sessions or meetings, and the like.
- **1997** The CSEE conference series broadened its focus to include training and known as the Conference on Software Engineering Education and Training (CSEET).
- **1997** Distance learning became a popular concept as well as the establishment of branch locations among universities.
- **1999** Establishment of the Software Engineering Code of Ethics and Professional Practice (IEEE-CS) and an associated curriculum.

**Present**

- The CSEET has evolved to include the Academy for Software Engineering Educators and Trainers (ASEET).
- New degree (PhD programs) in software engineering exist at a number of universities.
- New degree programs, such as the MS in Information Technology – Software Engineering Management (MSIT-SEM) continue to be defined.
- The number of conferences and conference tracks also has grown. Among of them are: Special Interest Group on Computer Science Education (SIGCSE, 2008), Innovation and Technology in Computer Science Education (ITiCSE), International Conference on Software Engineering (ICSE).

**Future**

- Maybe a Master's curriculum will initiated which includes Integrated Software and Systems Engineering Curriculum.
- Maybe three types of degree programs will initiated: computer science for researchers, software engineering for engineers developing mission–critical systems, and software professional degrees for practitioners developing non-critical systems.
- From the view point of globalization, a creative evolution of delivery mechanisms and a multicultural teams and offshoring (where you are located) will be expected.
- Conferences, working groups, and committees will help to develop a new generation of software engineering educators.
- e-publishing, video conferences, Webcasts and Web-based collaboration mechanisms are becoming more common.
- Professionalism continues to be a hot topic, and it's still unpredictable whether licensing or certification will become a major trend in the future. Litigation over software quality, safety, and security will become more common.
- New specializations will emerge.

Fig. 3. History of software engineering education: Past, present and future.

Multimedia represent an integrated system, enabling contact with users, taking their attention and offering the possibility for informing them [12]. They utilize technical means for simultaneous acting through visual tools, enabling easy and intuitive control through user friendly graphical interface. Key elements of the multimedia include static and animated graphics, sound and text, video sequences.

Software engineering faculty face the challenge of educating future researchers and industry practitioners regarding the generation of empirical software engineering studies and their use in evidence-based software engineering [17]. Professional software engineers are constantly faced with having to cope with ever-changing technologies, along with the need to keep their knowledge up to date. These changes, the short innovation cycles, and the fact that software engineering is a knowledge-intensive activity lead to many learning situations where new knowledge is required to solve the challenges and problems at hand. Furthermore, in practice learning, is less a reaction to 'being learned' but more the reaction to a variety of working situations and related problem-solving activities, which requires experience-based learning. Today, most software engineers are from the Baby Boomers generation (born 1946–1964) and from Generation X (born 1965–1980). A Net Generation (born 1981–1994) differs from the previous generations in terms of commitment, interaction, and learning style and hence puts new challenges upon education and the usage of technology [18, 22].

It is almost universally accepted that the best way to convey these ideas is through a laboratory or a demonstration, where students can see Metrology in action and truly appreciate the natural world around them [3, 23]. Figure 4 show the use of computer and Web-based engineering education and training.

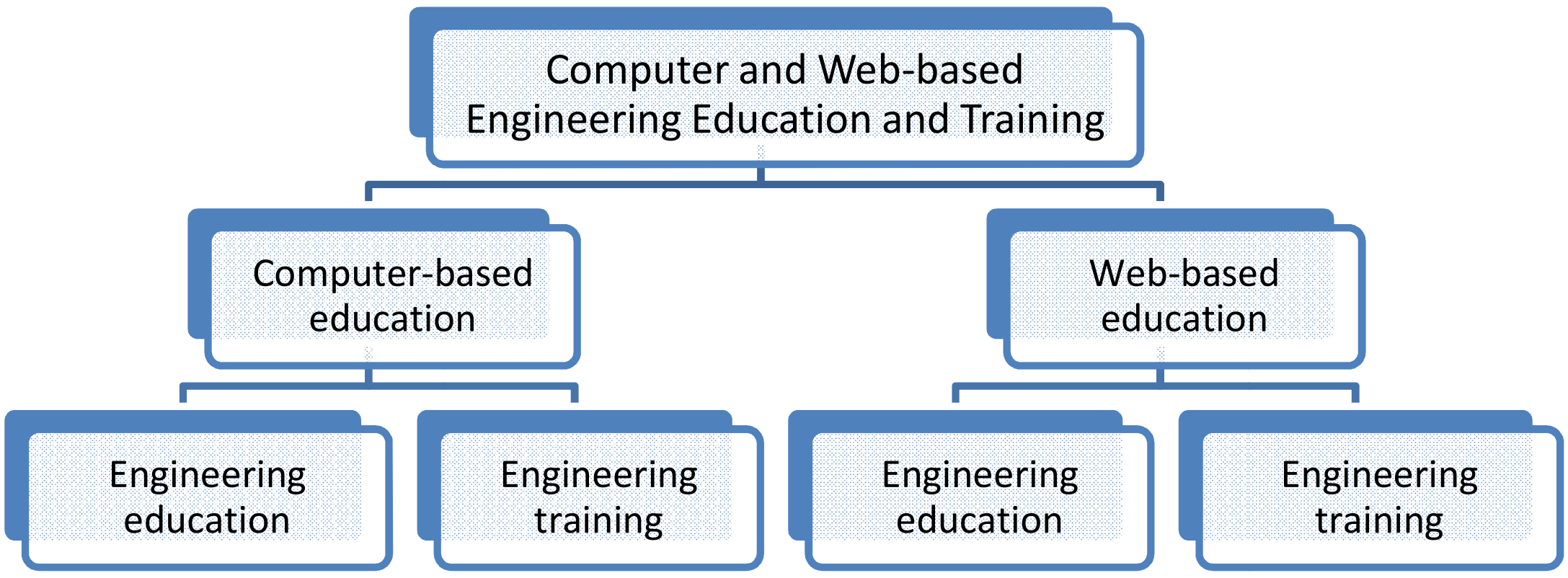

Fig. 4. Computer and Web-based engineering education.

## 3 HARDWARE AND SOFTWARE REQUIREMENTSs

The two essential requirements of Web-based education are hardware and main programming medium. The virtual labs can hosted on any computer connected to the internet; as most universities already have some internet-ready computers, such labs require no additional investment in hardware. The only requirements are a Java enabled browser. The applets that form the core of the computer-based laboratory experience use standard Java. These applets are accessible from any computer that has a browser and an internet connection. For those labs that lack an internet connection, it is possible to package the applets themselves, along with the associated lab, onto a CD and distribute these to the students. These mini applications provide total interactivity, combined with full multimedia and graphics that allow students to easily visualize difficult concepts. Incorporation of Java and Shockwave applets allow each student to have an interactive, hands-on experience in a lab, a personalized learning experience, to eliminate equipment issues, to minimize experimental errors, and to help making online learning resources not only much easier but also more fun for students. The main costs associated with the development of labs are programmer-time in setting up the hardware and software and instructor time in customizing the labs for each specific course.

## 4 CREATING WEB-BASED EXPERIMENTS

Many engineering universities are currently under pressure to develop Web-based online education and training via virtual laboratories and classrooms [24, 25]. In developing Web-based laboratories, four experiments were selected which have a clear learning value, that could easily be simulated, and whose results could be attractively displayed over the WWW. These experiments are Vernier caliper, micrometer, dial indicator and protractor widespread applications in length and angle measurements. The main objective is to develop a home page for the metrology and measurement course to be hosted on internet for engineering students. The user interface of these application is simply an HTML page that can be used to invoke a Java applets and multimedia. Several HTML pages which include explanation of engineering metrological concepts have been designed. The selected dimensional measurements have been included as PDF files and multimedia**.** More developments is still under construction mainly in the form of Java applets to be added to virtual metrology lab. Figures 5 shows the main page of the lab built using HTML programming. Figure 6 show the safety rules of metrology labs as a good true practice. These safety rules includes: Laboratory safety, emergency response, personal and general laboratory safety, electrical safety, mechanical safety, chemical safety, and additional safety guidelines.

## 5 FOUR SAMPLE LABS

The Web's impact on traditional educational theories and practices are increasingly apparent. New innovations such as virtual colleges, laboratories, and universities are developed. These include innovative hardware and software technology, online testing and assessment, training and teaching applications, and courseware design and development.

Metrology labs have been designed to lead the student and thereby develop his sense of enquiry into how things work. Metrology is not just a process of measurement that is applied to an end product. It should also be one of the considerations taken into account at the design stage. According to the geometrical product specification model, tolerancing and uncertainty issues should be taken into account during all stages of design, manufacture and testing.

The main goal is to show how it can be easy to understand metrology and to stimulate tools to do more exploration at home on their own computers. A set of computer-based metrology labs make use of Java applets that are available online or offline browsing. The provided labs were developed for the educational use of Albaha University. Students are introduced to basic measurement techniques such as length and angle measurements with vernier caliper, micrometer, dial indicator and protractor. The following shows lab user guide to use and to measure dimensions of objects using, for example, a Vernier caliper, Fig. 7.

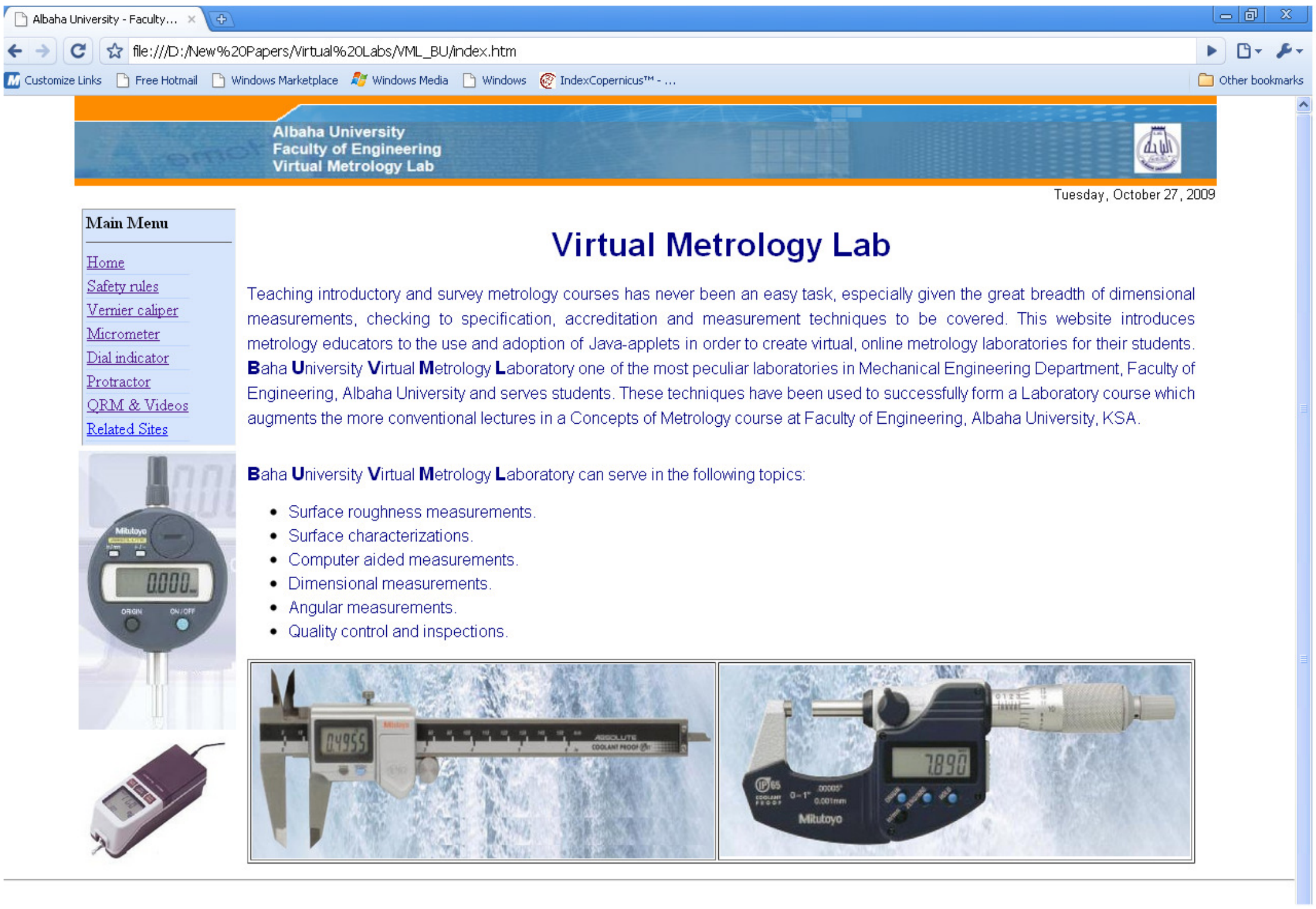

*Fig. 5. Main Webpage of Virtual Metrology Lab.*

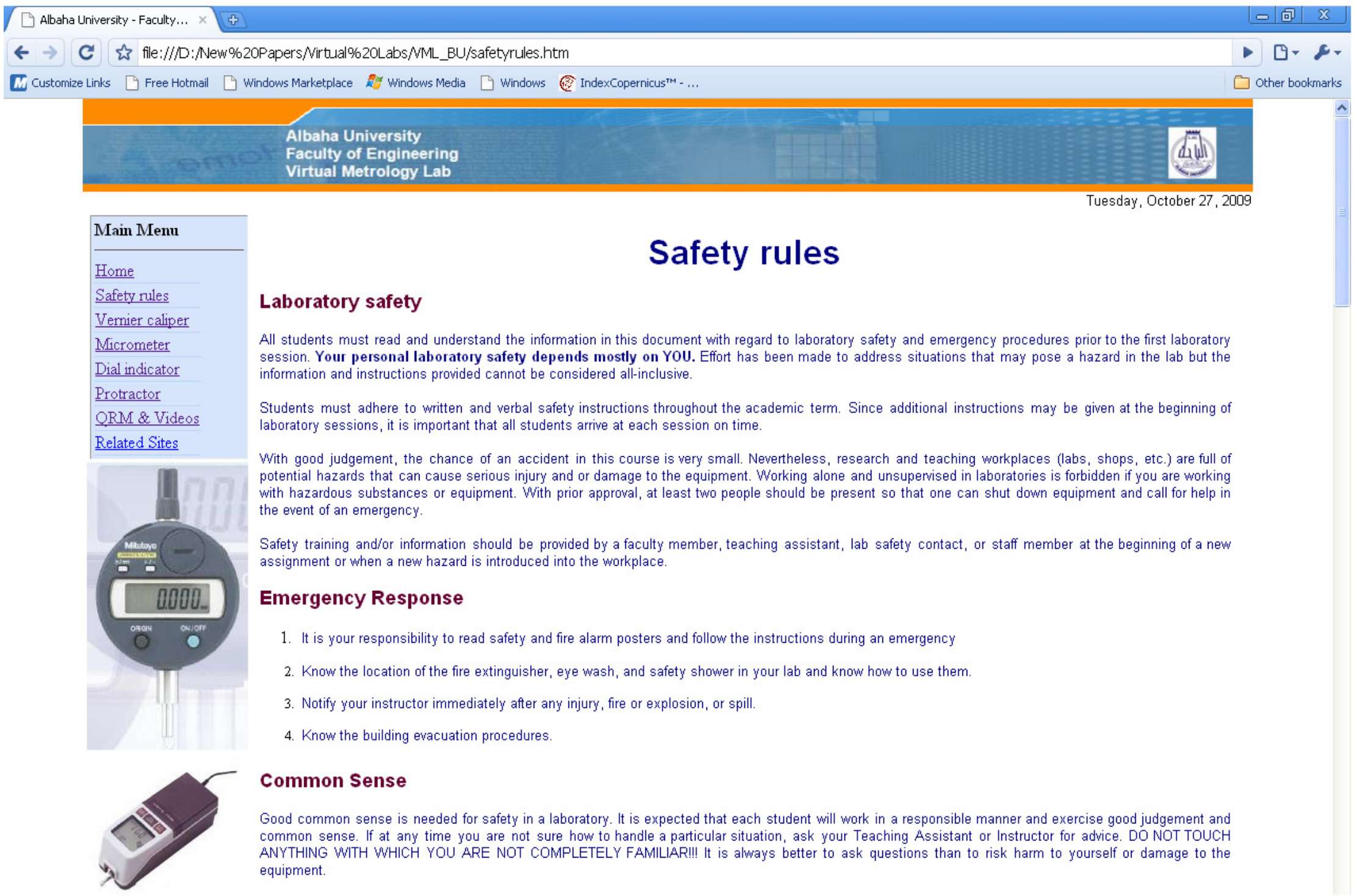

Fig. 6. Safety rules webpage of Virtual Metrology Lab.

The Java applet shows you how to read the vernier scale on Vernier calipers. The reading appear into the *Text Field* in mm. To observe how the scales are read, click in the *"checkbox"* by *"show reading"* and drag the lower scale left or right to observe how the measurement is determined. Notice that the value for the upper scale is determined by mm where the 0 on the lower scale lines up with the upper scale and the number on the lower scale is determined by finding which mark on the lower scale best lines up with any mark on the upper scale. Use the *"Reset"* button to restore the applet to the initial state. In order to test your ability to read a measurement, do not press "show reading" and then drag the lower scale left or right. The applet will wait for you to enter a value into the *Text Field*. If your answer is right, it will appear massage *"Well done"*. If your answer is not right, it will appear massage *"Sorry, wrong answer !"*. After you type in an answer, press *Enter*. Figure 8 shows a typical reading example for Vernier calipr, Micrometer, Dial indicator and Protractor.

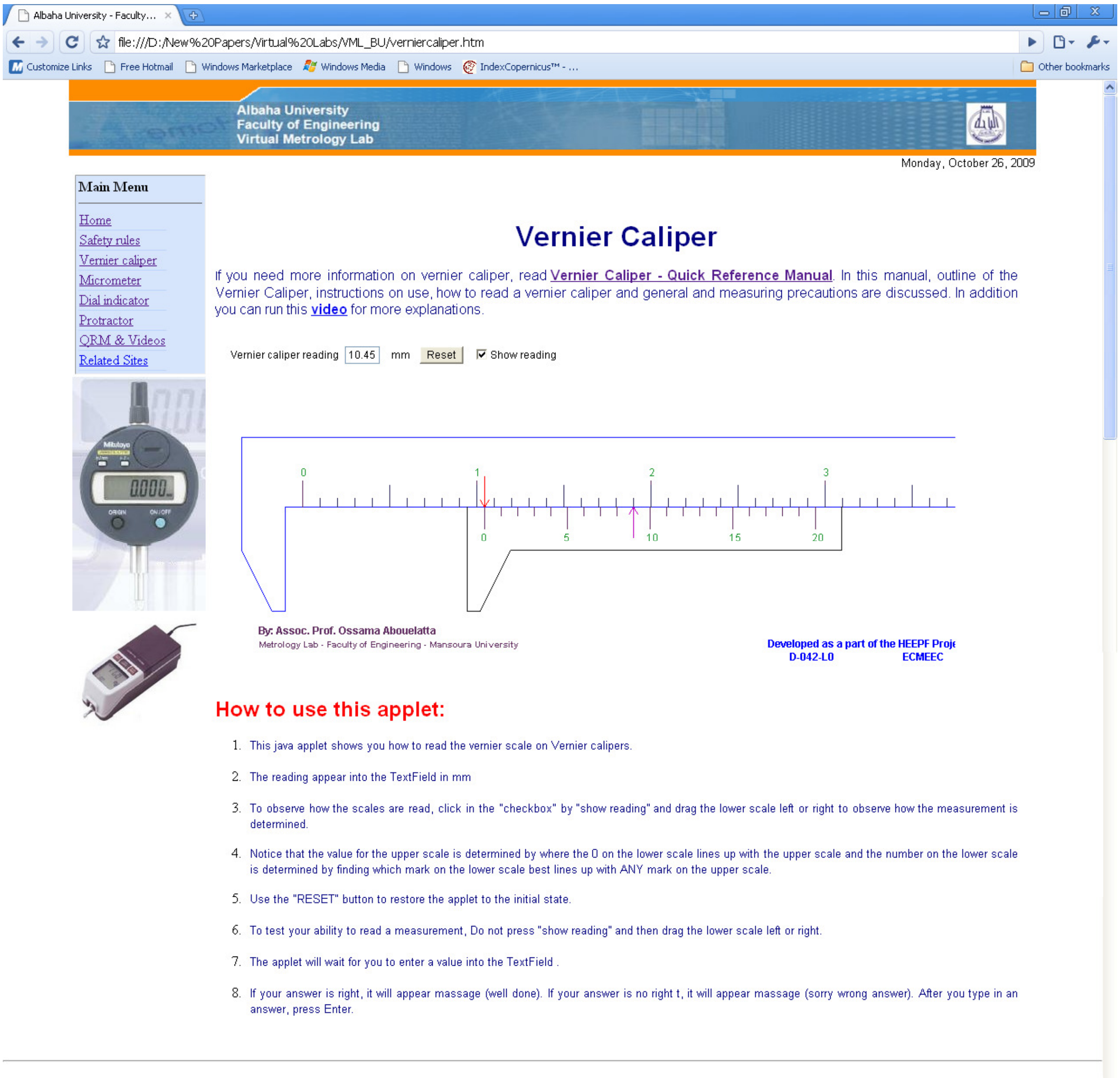

Fig. 7. Vernier caliper webpage of Virtual Metrology Lab.

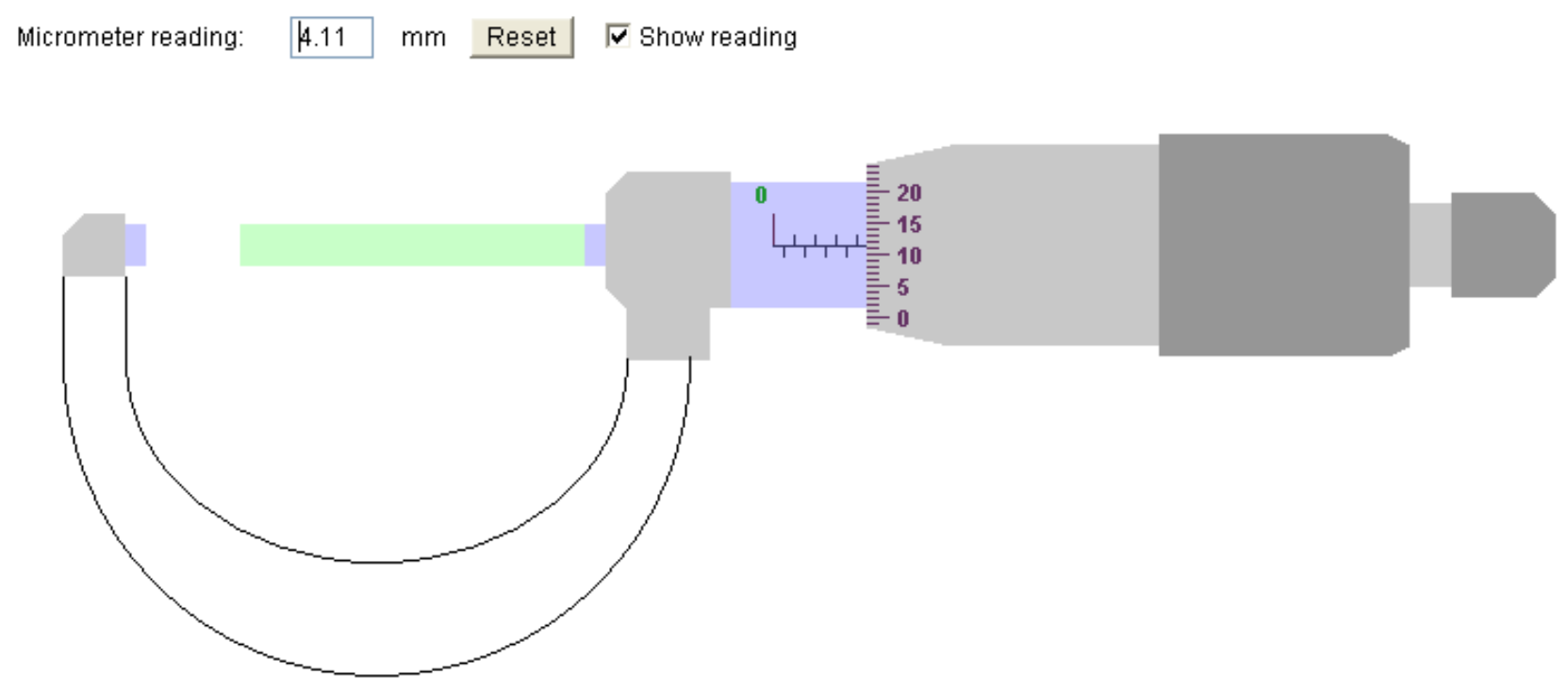

(a). Java Applet of Micrometer.

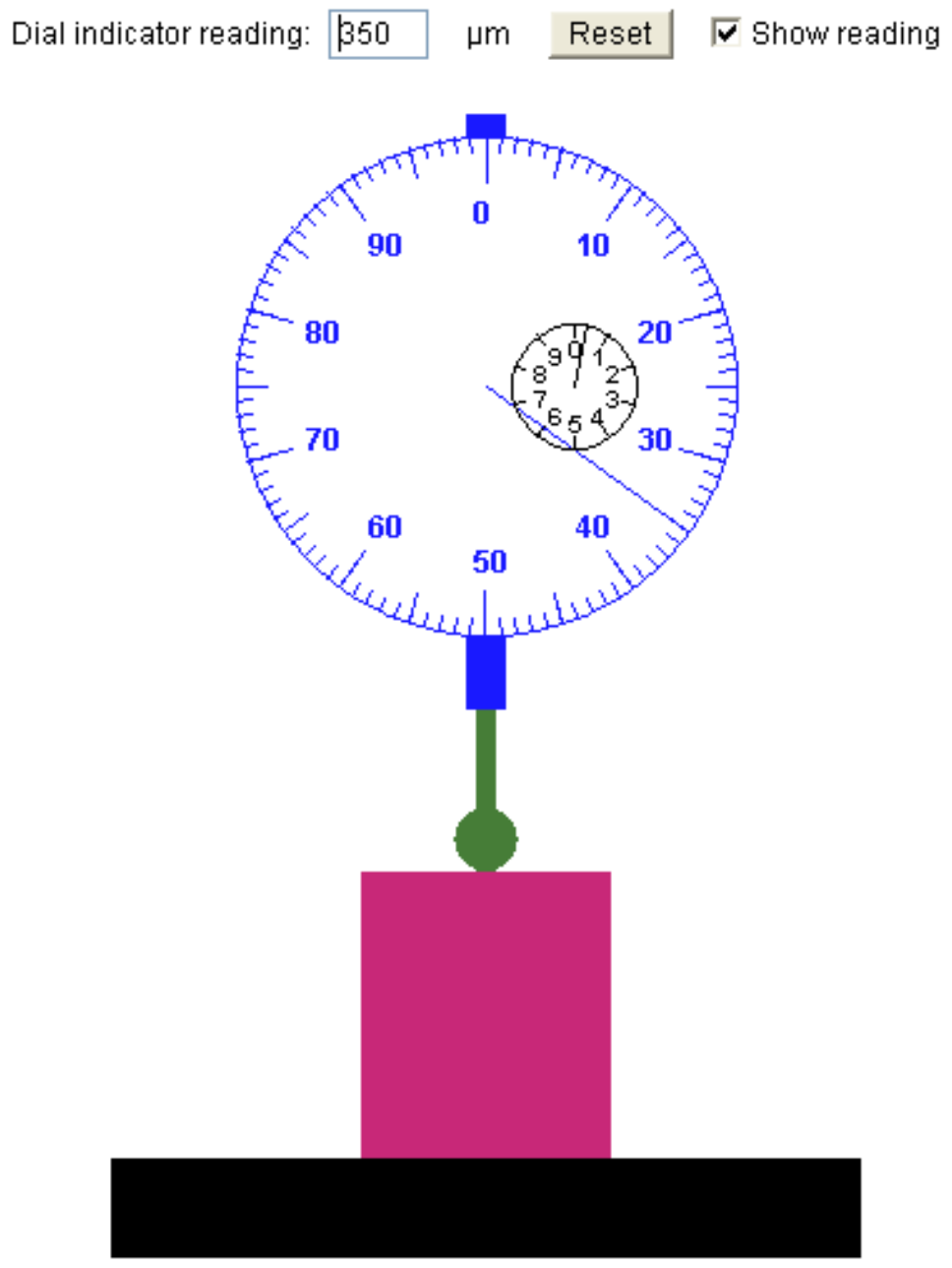

(b). Java Applet of Dial indicator.

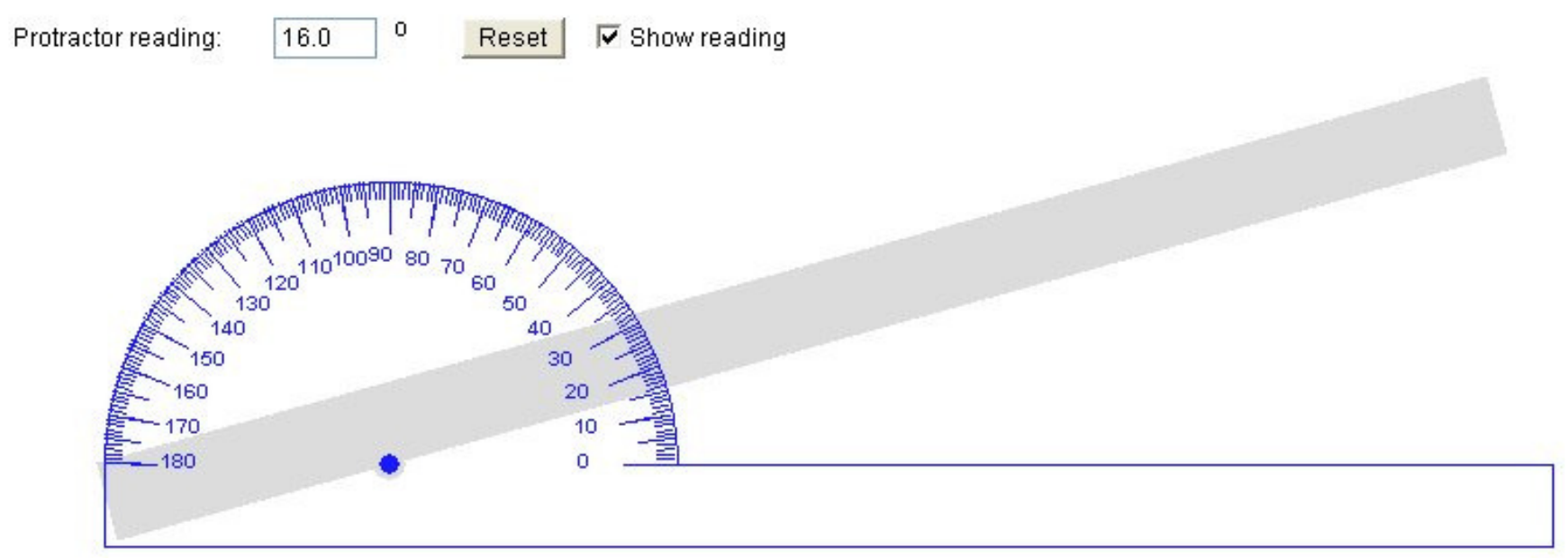

(c). Java Applet of Protractor.

Fig. 8. A typical reading example of (a) Micrometer, (b) Dial indicator, and (c) Protractor

## EVALUATION

Educational software firm was based on the vision that every student should be able to achieve personal success in reading and writing. The goal of instructional designers is to make learning easier, quicker, and more enjoyable. One of the main purposes of instructional designer's job is to help everyone to learn and be successful. So, it should be expected after reading educational materials, users will be able to make better measurements of the size or shape of an object concerning the introduced subject "Dimensional Metrology". The content is written at a simple way based on standard textbooks so that it can easily and quickly introduce key ideas to a wide audience. A preliminary survey analysis based on student's feedback was reported. This survey analysis helps in designing and modifying the strategy of off-line material to be published on-line.

## SUMMARY AND CONCLUSIONS

It is important to note that the aim of Virtual metrology labs is to provide enhanced educational and training support to the students in addition to traditional teaching and learning. It should be noted that this tool is very effective to increase the efficiency of learning and training. In a traditional university environment, virtual labs can also improve the engineering curricula in a cost effective way by establishing a timely connection between theory and practice [q8]. Web-based education and training offer many advantages to engineering and science education. Four Java applets: Vernier caliper, micrometer, dial indicator and protractor have been off-line run successfully with minimal problems. This experience encourages us to develop other web-based education and training, improving the site with more live video and Java applets. This Web-based education and training has been successfully class-tested within an undergraduate preliminary year engineering course and students reported a positive experience with its use. The use of these labs should be self-explanatory and their reliable operation has been thoroughly tested. In a future work, the usability, efficiency and effectiveness of the system both from the point of view of the lecturers and the students will be reported.

## REFERENCES

1. J. Ángel Velázquez-Iturbide, Francisco José García-Peñalvo, Computers in Education: Advances in Software Technology, Journal of Universal Computer Science, vol. 15, no. 7 (2009), 1423-1426

2. Nancy R. Mead, Software engineering education: How far we've come and how far we have to go, The Journal of Systems and Software 82 (2009) 571–575

3. Ricky J. Sethi, Using virtual laboratories and online instruction to enhance physics education, J. Phys. Tchr. Educ. Online 2(4), February 2005 Page 23-27.

4. Ricky J. Sethi, Gault A. Antcliffe, On-Line, Interactive Computer-Based Physics Laboratory, ELabs: Online Interactive Physics, http://www.sethi.org/classes/elabs/.

5. Bekim Fetaji, Majlinda Fetaji, Software Engineering Java Educational Software and its Qualitative Research, Current Developments in Technology-Assisted Education (2006), pp. 2057-2064

6. Pniower, J. C., Ruane, M., Goldberg, B. B. and Ünü, M. S. "Web-Based Educational Experiments", Proceedings of the 1999 ASEE National Conference, Charlotte, NC, June 1999, Session 3232.

7. Syropoulos A., Protopapas A., Greece, Web-Based Educational Use of Hydraulic Software in Fortran, XXX IAHR Congress, August 2003, Thessaloniki, Greece, pp. 179-184

8. Sheikh, Sharif Iqbal, Junaid Siddiqui, Web Based Engineering Education, The 6th Saudi Engineering Conference, KFUPM, Dhahran, December 2002, Vol. 1. Pp. 315-324

9. Philippos Pouyioutas, Maria Poveda, Dmitri Apraksin, The Impact of Web-Based Educational Software: Off-the-Shelf vs. In-House Developed Software, Journal of Information Technology Impact, Vol. 3, No. 3, pp. 121-130, 2003

10. Mohamed Aichouni, Mohamed Othmane Al Nais, Interactive demonstrations of statistical quality control for engineering students using computer based tools, The third forum on engineering education, sharjah university, U.A.E., October 14-15, 2003, pp. 1-6.

11. Arif Sirinterlikci, Active Learning in Manufacturing Engineering Programs, Proceedings of The 2008 IAJC-IJME International Conference, Paper 16, ENG 107

12. Martin Halaj, Peter Gabko, Eva Kurekova, Rudolf Palenčár, Project for the modern educational tool in measurement and Metrology, Measurement Science Review, Volume 3, Section 1, 2003, pp. 23-26

13. Jones, D., Gregor, S., & Lynch, T. (2003). An Information Systems Design Theory for Web-Based Education. IASTED International Symposium on Web-based Education, Rhodes, Greece, ACTA Press.

14. Ivan Milentijevic, Vladimir Ciric, Oliver Vojinovic, Version control in project-based learning, Computers & Education 50 (2008) 1331–1338

15. Panu Silvasti, Lauri Malmi, Petteri Torvinen, Collecting statistical data of the usage of a Web based educational software, Proceeding of the IASTED International Conference Web-Based Education, February 16-18, 2004, Innsbruck, Austria, pp. 107-110

16. Titus K.L. Schleyer, Lynn A. Johnson, Evaluation of Educational Software, Journal of Dental Education, November 2003, pp. 1221-1228

17. David S. Janzen, Jungwoo Ryoo, Engaging the net generation with evidence-based software engineering through a community-driven Web database, The Journal of Systems and Software 82 (2009) 563–570

18. Eric Ras, Jörg Rech, Using Wikis to support the Net Generation in improving knowledge acquisition in capstone projects, The Journal of Systems and Software 82 (2009) 553–562

19. Nigel Ford, Web-Based Learning through Educational Informatics: Information Science Meets Educational Computing, Information Science Publishing, Hershey, New York, 2008

20. Bruce L. Mann, Selected Styles in Web-Based Educational Research, Information Science Publishing, Hershey, New York, 2006

21. George D. Magoulas and Sherry Y. Che, Advances in Web-Based Education: Personalized Learning Environments, Information Science Publishing, Hershey, New York, 2006

22. Prensky, M., 2001. Digital natives, digital immigrants. On the Horizon 9 (5).

23. Mehdi Khosrow-Pour, Web-Based Instructional Learning, IRM Press, USA, 2002

24. Joseph T. Sinclair, LaniW. Sinclair and Joseph G. Lansing, Web-Based Training: A Step by Step Guide to Designing Effective E-Learning, Library of Congress Cataloging-in-Publication Data, USA, 2002

25. Chi Chung Ko and Chang Dong Cheng, Interactive Web-Based Virtual Reality with Java 3D, Information Science Publishing, Hershey, New York, 2009